\documentclass{llncs}

\usepackage{multirow}
\usepackage{subcaption}
\usepackage{comment}
\usepackage{amssymb}
\usepackage{graphicx}
\usepackage{amsmath}
\usepackage{url}
\usepackage{float}
\usepackage{pifont}
\usepackage{soul}
\usepackage{adjustbox}

\usepackage{silence}
\usepackage{xcolor}

\usepackage{soul}  
\setstcolor{red} 

\urldef{\mailsa}\path|mucahidkutlu@qu.edu.qa| 
\urldef{\mailsb}\path| vivek.khetank@utexas.edu|
\urldef{\mailsc}\path|ml@utexas.edu|

\begin{document}

\title{Correlation and Prediction of  Evaluation Metrics in Information Retrieval}

\author{Mucahid Kutlu\inst{1}, Vivek Khetan\inst{2}, Matthew Lease\inst{2}}

\institute{Computer Science and Engineering Department, Qatar University, Doha, Qatar\\
\and School of Information, University of Texas at Austin, USA\\
\mailsa, \mailsb, \mailsc}



\maketitle

\begin{abstract}

Because researchers typically do not have the time or space to present more than a few evaluation metrics in any published study, it can be difficult to assess relative effectiveness of prior methods for unreported metrics when baselining a new method or conducting a systematic meta-review. While sharing of study data would help alleviate this, recent attempts to encourage consistent sharing have been largely unsuccessful. Instead, we propose to enable relative comparisons with prior work across arbitrary metrics by predicting unreported metrics given one or more reported metrics. In addition, we further investigate prediction of high-cost evaluation measures using low-cost measures as a potential strategy for reducing evaluation cost. We begin by assessing the correlation between 23 IR metrics using 8 TREC test collections. Measuring prediction error {\em wrt.} $R^2$ and Kendall's $\tau$, we show that accurate prediction of MAP, P@10, and RBP can be achieved using only 2-3 other metrics. With regard to lowering evaluation cost, we show that RBP(p=0.95) can be predicted with high accuracy using measures with only evaluation depth of 30.
Taken together, our findings provide a valuable proof-of-concept which we expect to spur follow-on work by others in proposing more sophisticated models for metric prediction.
\end{abstract}


\section{Introduction}
\label{section:introduction}

To assess an IR system's effectiveness for different  search scenarios, researchers have proposed a wide variety of evaluation metrics, each providing a different view of system effectiveness~\cite{aslam2005maximum}.  
For example, while \emph{precision@10} (P@10) and \emph{reciprocal rank} (RR) are often used to evaluate the quality of the top search results, \emph{mean average precision} (MAP) and \emph{rank-biased precision} (RBP)~\cite{moffat2008rank} are often used to quality of search results at greater depth.

Popular evaluation tools such as {\tt trec\_eval}\footnote{\url{trec.nist.gov/trec_eval/}}  compute many more evaluation metrics than IR researchers typically have time or space to analyze and report. Even for the most knowledgeable and diligent researcher, it is challenging to decide which small subset of metrics should be reported to best characterize a given IR system's performance. Of course, presenting only a few metrics cannot fully characterize system performance. Information is thus lost in publication, and some interested reader will be disappointed to find a particular desired metric missing, especially when trying to baseline a new method for a given metric, or when conducting a meta-review comparison of prior work.

To compute a different metric of interest, one strategy is to try to reproduce prior work. However, this is often difficult (and sometimes impossible) in practice, as the written description of a method is often incomplete and even shared sourcecode can be difficult or impossible for others to run, especially as compilers, programming languages, and operating systems change. Another strategy is to share system outputs, enabling others to compute any metric of interest for those outputs. While Armstrong et al.~\cite{armstrong2009improvements} proposed and deployed  a central repository\footnote{\url{www.evaluatIR.org}} to store IR system runs, their proposal did not achieve broad buy-in from IR researchers and was ultimately abandoned. Realistically, it seems such broad buy-in is unlikely unless eventually mandated by research funding agencies. A similar situation exists in the field of biomedical literature mining \cite{hirschman2002accomplishments,de2002literature}, where lack of shared data has generated a large body of research in mining published papers to infer additional information. With published papers being the most standard and enduring record of research studies, the capacity to predict an arbitrary metric of interest given only one or more other metric scores, easily obtained from published studies, could be quite valuable in practice.



Another potential application of such prediction could be to decrease the massive cost of evaluation by enabling prediction of high-cost measures using low-cost measures. That is, instead of collecting many relevance judgments to calculate a particular high-cost measure (e.g. MAP@1000), we would rather collect fewer judgments,  calculate any number of low-cost measures (e.g. P@10, MAP@10, nDCG@10) and predict a high-cost measure of interest. 

To address this challenge, we first investigate the correlation between a wide range of evaluation metrics. Using runs submitted to 8 TREC tracks, we compute  23 evaluation measures for every track, system, and topic in order to assemble a large database of paired metric scores. We then calculate Pearson correlation between each evaluation measure pairs. In our extensive experiments, we find out that many metrics are strongly correlated (i.e., $\rho>0.9$) such as:
\begin{itemize}
\item \emph{average precision} (AP),  \emph{R-Precision} (R-Prec), and bpref
\item RBP(p=0.5) and RR 
\item RBP(p=0.95), RBP(p=0.8), P@10 and P@20
\item nDCG@20 and RBP(0.8).
\end{itemize}
%
%
Following this, we report use of linear regression to predict one  metric given 1-3 other metrics. We explore prediction of 12 measures and evaluate our prediction model on 3 test collections. Results show we can accurately predict:
\begin{itemize}
\item MAP given nDCG and R-Prec
\item P@10 given RBP(p=0.5)  and RBP(p=0.8)
\item RBP(p=0.5) given RR and RBP(0.8)
\item RBP(p=0.8) given P@10, RBP(p=0.5) and RBP(p=0.95)
\end{itemize}
%
Therefore, if a system's performance is reported  with these measures, we can still reliably predict its performance on the respective measure. 

Finally, we investigate prediction of high-cost measures using low-cost measures. We show we can accurately predict  RBP(p=0.95) at evaluation depth of 1000 and 100 given  measures computed at depth 30, which shows the promise of this strategy for lowering evaluation cost.

Contributions of our work include:
\begin{itemize}
\item We analyze correlation between 23 metrics, using more recent collections than prior work. This includes \emph{expected reciprocal rank} (ERR) and RBP using graded relevance judgments, whereas relevant prior work used only binary relevance judgments for these metrics.

\item We show that accurate prediction of metrics can be achieved  using only 2-3 other metrics. Further improvements can be expected using more sophisticated prediction models and larger training data.

\item We show that our prediction model can also be used to decrease the cost of evaluation by predicting  high-cost measures using low-cost measures. 
\end{itemize}

Section~\ref{section:related-work} discusses the prior  work. Section~\ref{section:data} describes the data used in our experiments.
Section~\ref{section:correlation}  and~\ref{section:prediction} present correlation and prediction of  evaluation metrics, respectively. Finally, we conclude in Section~\ref{section:conclusion}.

\section{Related Work}
\label{section:related-work}

In order to  better understand similarity between evaluation metrics, several studies have investigated correlation between  them.

Tague-Sutcliffe and Blustein~\cite{sutcliffe95} investigate correlation between 7 measures on TREC-3 data and show that R-Prec and AP are strongly correlated. This high correlation between R-Prec and MAP is also confirmed by Buckley and Voorhees~\cite{buckley2005retrieval} using Kendall's $\tau$ on TREC-7. 
Baccini et al.~\cite{baccini2012many} measure correlations between 130 measures calculated by {\tt trec\_eval} using data from the TREC-(2-8) ad hoc task and group them into 7 clusters based on correlation.  Sakai~\cite{sakai2007reliability} compares 14 graded-level and 10 binary level metrics  using three different data sets from NTCIR. In another work~\cite{sakai2007properties}, Sakai studies correlation between P($^+$)-measure, O-measure, normalized weighted reciprocal rank and RR, and concludes that 
they are highly correlated each other except RR. Egghe~\cite{Egghe2008856} investigates the correlation between precision, recall, fallout and miss.  Ishioka~\cite{ishioka2003evaluation} explores relation between  F-measure, break-even point, and 11-point averaged precision. 
Thom et al.~\cite{thom2007comparison} also studies correlation between 5 evaluation measures using TREC Terabyte Track 2006.
None of these works cover ERR and RBP; we investigate correlation of 23 measures including ERR and RBP.

Jones et al.~\cite{jones2015features} examine disagreement between 14 evaluation metrics including ERR and RBP  using TREC-(4-8) ad hoc tasks, and  TREC Robust 2005 and 2006 tracks. However, they use only binary relevance judgments in their analysis, which makes ERR identical to RR, whereas we consider graded relevance judgments. 
In addition, the most recent test collections used in this related prior work is TREC Robust Track 2006 and Terabyte Track 2006. In contrast, we consider more recent TREC test collections (i.e. Web Tracks 2010-2014).

A primary contribution of our work is investigating prediction of evaluation measures. While Aslam et al.~\cite{aslam2005geometric} also proposes predicting evaluation measures, they require a corresponding retrieved ranked list as well as another evaluation metric. They conclude that they can infer accurately user-oriented measures (e.g.\ P@10) from system-oriented measures (e.g.\ AP, R-Prec). In contrast, we predict evaluation measure of a system given only other evaluation measures  without requiring the corresponding ranked lists.

\section{Experimental Data}\label{section:data}
In order to investigate correlation and prediction of evaluation measures, we used the submitted runs and relevance judgments of Web Tracks (WT) of TREC-2000, 2010-2014 and Robust Track (RT) of TREC-2004. We consider only {\em ad hoc} retrieval. \textbf{Table~\ref{data}} lists the test collections used in our study. 

\begin{table}[H]
\centering
\caption{TREC Test Collections Used in Our Study.}
\begin{tabular}{|c|p{4cm}|c|p{1.5cm}|}
\hline
      \bf Test Collection           		  & \bf Document Collection & \bf \# Systems & \bf Topics \\ \hline \hline
WT2000~\cite{hawking2000overview}		 &	WT10g		&    105   	&  451-500\\ \hline
WT2001~\cite{voorhees2001overview}      &	WT10g		&  97     	& 501-550 \\ \hline
RT2004~\cite{voorhees2004overview}   &	TREC disks 4\&5, minus the Congressional Record		&     110  	&  301-450, 601-700\\ \hline
WT2010~\cite{trec2010}      & ClueWeb'09		&  55     	& 51-99 \\ \hline
WT2011~\cite{trec2011}     &	ClueWeb'09		&  62     	& 101-150 \\ \hline
WT2012~\cite{trec2012}      &	ClueWeb'09		&  48     	& 151-200 \\ \hline
WT2013~\cite{trec2013}      &	ClueWeb'12		&  59     	& 201-250 \\ \hline
WT2014~\cite{trec2014}      &	ClueWeb'12		&  30     	& 251-300 \\ \hline
\end{tabular}
\label{data}
\end{table}

Using the system runs submitted to these selected TREC tracks and their respective relevance judgments, we calculated 9 different evaluation metrics, including AP,  bpref~\cite{buckley2004retrieval}, ERR~\cite{chapelle2009expected}, nDCG, P@K,  RBP~\cite{moffat2008rank}, \emph{recall} (R), RR~\cite{voorhees1999trec}, and R-Prec. 
We used various cut-off thresholds for the metrics. The cut-off threshold for a particular metric is shown by "@" sign followed by the threshold value (e.g. P@10, R@100). Unless stated, we set the cut-off threshold to 1000, which is {\tt trec\_eval}'s default.
The cut-off threshold for ERR is set to 20 because it has been used as one of the official measures in WT2014. 
RBP  uses a parameter, called \emph{p}, representing the probability of a user desiring to see the next retrieved page. 
In our calculations, we test  0.5, 0.8 and 0.95 for the $p$ parameter, which are also the $p$ values explored by Moffat and Zobel~\cite{moffat2008rank}. Using these metrics, we generated two datasets:

\begin{itemize}
\item \textbf{Topic-Wise (TW) Dataset:} We calculated each metric mentioned above for each system for each separate topic. We used 10, 20, 100 and 1000 cut-off thresholds for AP, nDCG, P@K and R@K. In total, we calculated 23 evaluation measures. 

\item \textbf{System-Wise (SW) Dataset:} We calculated each metric mentioned above for each system, averaging over all topics in the corresponding test collection. For AP score, in addition to MAP, we also calculated \emph{GMAP} (i.e.\ geometric mean of AP). 
\end{itemize}

In order to calculate RBP and ERR, we used the RBP implementation provided by its authors\footnote{http://people.eng.unimelb.edu.au/ammoffat/rbp\_eval-0.2.tar.gz} and the ERR implementation\footnote{https://github.com/trec-web/trec-web-2014} provided by TREC. For the rest of the performance measures, we used {\tt trec\_eval 9.0}. As in any large dataset, various runs had missing data that resulted in only a subset of evaluation measures being computed. In such cases, we filtered out any such suspicious null or zero values. We also detected runs that have identical ranked lists in WT2013 and WT2014 test collections and filtered out identical submissions. 

\section{Correlation of Measures}\label{section:correlation}
\begin{figure}[h]
\centerline{\includegraphics[width=1\textwidth]{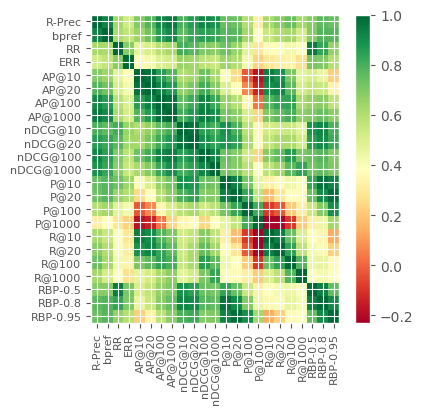}} 
\caption{Pearson Correlation between Metrics} 
\label{topic_wise_heatmap}
\end{figure}

 We studied the correlation of measures using the TW dataset instead of the SW dataset to avoid losing any information by averaging scores across topics. In particular, we  calculated Pearson correlation between measures across different topics using system runs in all test collections mentioned in Table~\ref{data}. 
The  correlation results are shown in \textbf{Figure~\ref{topic_wise_heatmap}}.

There are several observations we can make from these results. First, R-Prec has high correlation with bpref, MAP and nDCG@100, confirming prior work's findings that MAP and R-Prec are highly correlated~\cite{sutcliffe95,buckley2005retrieval,aslam2005geometric}. Second, RR is strongly correlated  with RBP(p=0.5) and its correlation with RBP measures decreases as the $p$ parameter of RBP increases. This is because as $p$ increases, RBP becomes more of a deep-rank metric while RR metric ignores the documents ranked after the first relevant document. 
Third, nDCG@20, which is used as one of the official metrics of WT2014, is  highly correlated with RBP(p=0.8). This finding indirectly verifies that nDCG@20 is an appropriate measure for web search tasks, connecting with Park and Zhang's~\cite{park2007distribution} suggestion that p=0.78 is an appropriate  value of RBP for modeling behaviour of web users. 
Fourth, nDCG is highly correlated with MAP and R-Prec and its correlation with R@K consistently increases as $K$ increases. 
Fifth, 
 most correlated  with RBP(p=0.8) and RBP(p=0.95) are P@10 ($\rho=0.97$) and P@20 ($\rho=0.98$), respectively. 
Sixth, Sakai and  Kando~\cite{Sakai2008} report that RBP(p=0.5) basically ignores relevant documents ranked lower than 10. Our results are consistent with this finding such that the maximum Pearson correlation between RBP(p=0.5) and  nDCG@K is obtained when K=10, and this correlation decreases as K increases.
Finally, among all measures, P@1000 is the least correlated one with others, suggesting that it captures an    effectiveness measure of IR systems that no other metric does.

\section{Prediction of  Metrics}\label{section:prediction}
In this section, we  describe our prediction model and experimental setup,  and report results of experiments we conducted to investigate prediction of evaluation measures. 

\subsection{Prediction Model \& Experimental Setup}\label{sec_model}
One key goal of our work is to predict a system's missing evaluation measure using reported ones. Thus,
we build a linear regression model  using only evaluation measures  of systems as features.  We use the SW dataset in our experiments for prediction because studies generally report their average performance over a set of topics, instead of reporting their performance for each topic.
We use data extracted from WT2000, WT2001, RT2004, WT2010 and WT2011 as the training dataset. WT2012, WT2013 and WT2014 are used to evaluate our prediction model. In order to evaluate the prediction accuracy, we report $R^2$ and  Kendall's $\tau$ correlation. 

\subsection{Prediction Using Varying Number of Measures}\label{section_predict_with_k_measures}

In this section, we explore the best predictors for 12  evaluation measures including  R-Prec, bpref, RR, ERR@20, MAP, GMAP, nDCG, P@10, R@100, RBP(0.5), RBP(0.8) and RBP(0.95).  Researchers can report different combinations of evaluation measures, yielding a huge number of cases we might consider. In order to reduce our search space, we investigate which $N$ evaluation measure(s) are the best predictors for a particular measure and vary N from 1 to 3. Specifically, in prediction of a particular measure, we try all combinations of size $N$ using the remaining 11 evaluation measures on WT2012 and pick the one that yields the best Kendall's $\tau$ correlation. Then, the selected combination of measures are used for predicting the respective measure on WT2013 and WT2014. The experimental results are shown in  \textbf{Table~\ref{system_3_labels}}.  Kendall's $\tau$ scores higher than 0.9 (a traditionally-accepted threshold for an acceptable correlation~\cite{voorhees2000variations}) are bolded. 

\textbf{bpref.} We achieve the highest $\tau$ correlation and interestingly the worst $R^2$ using only nDCG on WT2014. This shows that while predicted measures are not accurate, rankings of systems based on predicted scores can be highly correlated with the actual ranking. We observe the same pattern of results in prediction of RR on WT2012 and WT2014,  R-prec on WT2013 and WT2014,  R@100 on WT2013, and  nDCG in all three test collections. 

\textbf{GMAP \& ERR.} GMAP and ERR seem to be the most challenging measures to predict because we could never reach 0.9 $\tau$ correlation in any of the prediction cases of these two measures.  
 Initially, $R^2$ scores we achieve for ERR consistently increase in all three test collections as we use more evaluation measures for prediction, suggesting that we can achieve  higher prediction accuracy using more independent variables.

\textbf{MAP.} We can predict MAP with very high prediction accuracy and achieve higher than 0.9 $\tau$ correlation in all three test collections using R-Prec and nDCG as predictors. As we use RR as the third predictor, $R^2$ increases in all cases and $\tau$ correlation slightly increases on average (0.924 vs.\ 0.922).

\textbf{nDCG.} Interestingly, we achieve the highest $\tau$ correlations using only bpref; $\tau$   decreases as  more evaluation measures are used as independent variables.  Even though we reach high $\tau$ correlations for some cases (e.g. 0.915 $\tau$ on WT2014 using only bpref),   nDCG seems to be one of the hardest measures 
to predict. 

\textbf{P@10.} Using RBP(0.5) and RBP(0.8), which are both highly correlated measures with P@10, we are able to achieve very high $\tau$ correlation 
and   $R^2$ 
in all three test collections (0.912 $\tau$ and 0.983 $R^2$ on average). We reach nearly perfect prediction accuracy ($R^2=0.994$) on WT2012.

\textbf{RBP(0.5).} In all three prediction cases, RR is selected  as one of the independent variables, as expected because of being the most correlated measure with RBP(0.5) (See Figure~\ref{topic_wise_heatmap}). While using only RR is not sufficient to reach 0.9 $\tau$ correlation, when we use also RBP(0.8) (the second most correlated measure) we reach very high prediction accuracy in all three test collections (0.919 $\tau$ and 0.924 $R^2$ on average).

\textbf{RBP(0.8).} P@10 is the most correlated measure with RBP(0.8) and is selected as one of the independent variables in all cases, as expected. Using P@10 and RBP(0.5), we are able to achieve more than 0.9 $\tau$ correlation and more than $0.98$ $R^2$ in all test collections. Using  P@10, RBP(0.5) and RBP(0.95), we achieve the highest $R^2$ (0.998) and $\tau$(0.973) among all 108 cases (i.e., 3 test collections x  12 measures x 3 different independent variable sets).

\textbf{RBP(0.95).} Compared to RBP(0.5) and RBP(0.8), we achieve noticeably lower prediction performance, especially on WT2013 and WT2014. On WT2012, which is used as the development set in our experimental setup, we reach high prediction accuracy when we use 2-3 independent variables.

\textbf{R-Prec, RR and R@100.} In predicting these three measures, while we reach high prediction accuracy in many cases,  there is no independent variable group yielding high prediction performance on all three test collections.

Overall,  we achieve high predicion accuracy for MAP, P@10, RBP(0.5) and RBP(0.8) on all test collections. 
RR and RBP(0.8) are the most frequently selected independent variables (10 and 9 times, respectively). Generally, using a single measure is not sufficient to reach 0.9 $\tau$ correlation. However, we are able to achieve very high prediction accuracy using only 2 measures for many  scenarios.

\begin{table}
\begin{tabular}{|p{1.75cm}|c|c|c|c|c|c|c|c|c|}  \hline
 \multirow{2}{1.75cm}{\bf Predicted Metric} & \multicolumn{3}{c|}{\bf Independent Variables}  & \multicolumn{2}{c|} {\bf WT2012}  &  \multicolumn{2}{c|} {\bf WT2013}  & \multicolumn{2}{c|}{\bf WT2014}   \\ \cline{5-10}
 	& 	  \multicolumn{3}{c|}{}	& $\tau$& $R^2$ & $\tau$& $R^2$ & $\tau$ 	& $R^2$  \\ \hline \hline
 \multirow{3}{1cm}{bpref}	& nDCG 	& - 	 & -		&0.805		&-0.693	& 0.885	&0.079	&\bf 0.915	&	-1.174 \\ \cline{2-10}
 							& nDCG 	& R-Prec & - 		&0.872		& -0.202& 0.850	&0.094	&	0.824	&	-0.989  	\\ \cline{2-10}
  							& nDCG 	& R-Prec & R@100	&\bf 0.906 	& 0.284	& 0.844 & 0.645 & 0.866 	&	0.390   \\\hline\hline
\multirow{3}{1cm}{ERR}		& RR & - 	& -				&0.764		&-1.874	&0.734	&0.293	&0.704		&	-1.004 \\ \cline{2-10}
							& RR & RBP(0.8)		& - 	&0.790		&-1.809	&	0.777&0.392	&	0.714	&	-0.686 \\ \cline{2-10}
 							& RR& RBP(0.8) & R@100		&0.796 		& -1.728& 0.741 		& 0.478 & 0.704	 	&	-0.473  \\ \hline\hline
\multirow{3}{1cm}{GMAP} 	& bpref	& - 	& -			&0.729		&-1.216	&0.704	&-2.982	&	0.739	&	-1.034 	\\ \cline{2-10}
 							& nDCG & RBP(0.5)	& - 	& 0.817		&0.877	&0.777	&0.600	&	0.767	&	0.818  \\ \cline{2-10}
							& nDCG& RBP(0.95)  & RR 		&0.817	 	& 0.882 & 0.748 		& 0.514 & 0.794 	&	0.854 \\ \hline\hline
\multirow{3}{1cm}{MAP} 		& R-Prec & - 	& -			&0.885		&0.754	&0.824	&0.667	&\bf 0.952	&0.819 \\ \cline{2-10}
  							& R-Prec & nDCG 	& - 	& \bf 0.904	&0.894	&\bf 0.905&0.760&\bf 0.958	&	0.897  \\ \cline{2-10}
 							& R-Prec& nDCG  & RR		&\bf 0.924 	& 0.916 &\bf 0.901	& 0.779 &\bf 0.947 	&	0.922  \\ \hline\hline
\multirow{3}{1cm}{nDCG} 	& bpref& - 	& -				&0.805		&-2.101	&0.885	&-0.217	&\bf 0.915	&	-2.008		\\ \cline{2-10}
  							& bpref & GMAP		& - 	& 0.803		&-0.079	&0.809	&0.574	&	0.872	&	0.024  \\ \cline{2-10}
 							& bpref& GMAP& RBP(0.95) 	&0.794 		& -0.113& 0.801	 	& 0.556 & 0.850	 	&	-0.032 \\ \hline\hline
\multirow{3}{1cm}{P@10} 	& RBP(0.8)	& - 	& -		&0.884		&0.942	&0.832	&0.895	&0.866		&	0.893 	\\ \cline{2-10}
  						 	& RBP(0.8) & RBP(0.5) & - 	& \bf 0.941	&0.994	&0.882	&0.966	&\bf 0.914	&	0.988 	\\ \cline{2-10}
 							& RBP(0.8)& RBP(0.5) & RR 	&\bf 0.946	& 0.994	& 0.885	 	& 0.968 &\bf 0.914  &	0.987 \\ \hline\hline
\multirow{3}{1cm}{RBP(0.5)}	& RR	& - 	& -			&0.782		&0.901	&0.806	&0.921	&	0.810	&	0.878 	\\ \cline{2-10}
  							& RR & RBP(0.8)	& - 	 	& \bf 0.938	&0.935	&0.894	&0.934	& \bf 0.926	&	0.903 	\\ \cline{2-10}
 							& RR&  RBP(0.8) & nDCG	 	&\bf 0.936	& 0.916	& 0.882		&0.917	&\bf 0.942	&	0.885 \\ \hline\hline
\multirow{3}{1cm}{RBP(0.8)}	& P@10	& - 	& -			&0.884		&0.932	&0.832	&0.885	&	0.866	&	0.894  \\ \cline{2-10}
   							&  P@10 &RBP(0.5)	& - 	& \bf 0.963	&0.997	&\bf 0.919&0.986&\bf 0.947	&	0.992 	\\ \cline{2-10}
				 			& P@10& RBP(0.5)& RBP(0.95)	&\bf 0.973 	& 0.998	& \bf 0.916	&0.990	&\bf 0.968	&	0.997 \\ \hline\hline
\multirow{3}{1cm}{RBP(0.95)}& R-Prec& - 	& -			&0.824		&0.346	& 0.651	&-0.786	&	0.607	&	-2.401 	\\ \cline{2-10}
 							& bpref & P@10		& - 	& \bf 0.911	&0.952	&0.718	&0.873	&0.728		&	0.591 \\ \cline{2-10}
							& bpref& P@10& RBP(0.8)		&\bf 0.911	& 0.967	& 0.720	 	&0.868	& 0.744 	&	0.639\\ \hline\hline
\multirow{3}{1cm}{R-Prec}	& R@100 & - 	& -	&0.899	&0.708		& 0.871	&0.624	&\bf 0.935	&	0.019 \\ \cline{2-10}
 							& R@100 & RBP(0.95)	& - & \bf 0.909		& 0.952	&	0.820&0.882	&	0.820	&	0.759 \\ \cline{2-10}
 							& R@100 & RBP(0.95) & GMAP 	&\bf 0.924	& 0.970	& 0.833		& 0.914 & 0.841 	&	0.825  \\ \hline\hline
\multirow{3}{1cm}{RR} 		& RBP(0.5) & - 	& -			&0.782		&0.904	& 0.806	&0.927	&	0.810	&	0.878		\\ \cline{2-10}
							& RBP(0.5) & RBP(0.8)	& - 	&0.869		&0.918	&	0.809&0.919	&	0.820	&	0.942  	 \\ \cline{2-10}
							& RBP(0.5)& RBP(0.8) & ERR	&0.876  	& 0.437 & 0.818 		& 0.924	&\bf 0.915 	&	0.824  \\ \hline\hline
\multirow{3}{1cm}{R@100} 	& R-Prec& - 	& -			&0.899		&0.423	& 0.871	&0.232	&\bf 0.935	&-1.075 	\\ \cline{2-10}
 							& R-Prec & GMAP		& - 	&0.899		&0.433	&0.871	&0.238	&\bf 0.940	&	-1.077  \\ \cline{2-10}
  							& R-Prec& RR& ERR	 		&0.881	 	& -0.104 & 0.823 	& 0.355 &\bf 0.935	&	-1.187 \\ \hline
\end{tabular}
\caption{System-wise Prediction Using Varying Number of Metrics. Kendall's $\tau$ scores higher than 0.9 are bolded.}
\label{system_3_labels}
\end{table}



\subsection{Prediction of High-Cost Measures with Low-Cost Measures}

\begin{figure}
    \centering
    \begin{subfigure}[b]{\textwidth}
        \includegraphics[width=\textwidth]{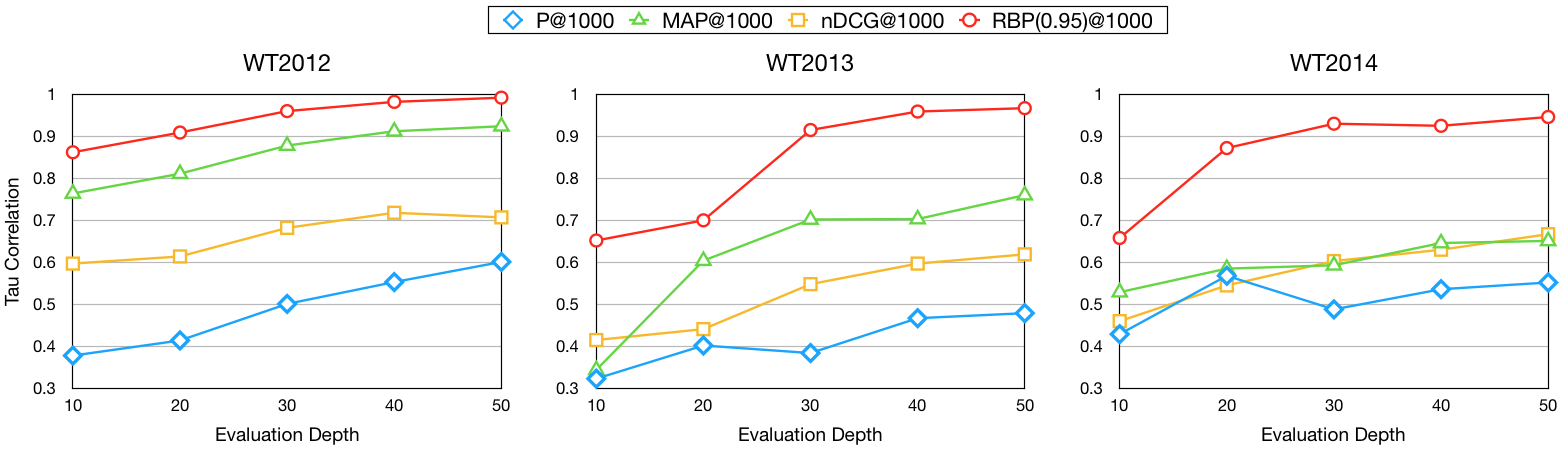}
        \caption{Judgment Depth of High-Cost Measures is 1000}
        \label{figure_low_to_high_cost_1000}
    \end{subfigure}
    \begin{subfigure}[b]{\textwidth}
        \includegraphics[width=\textwidth]{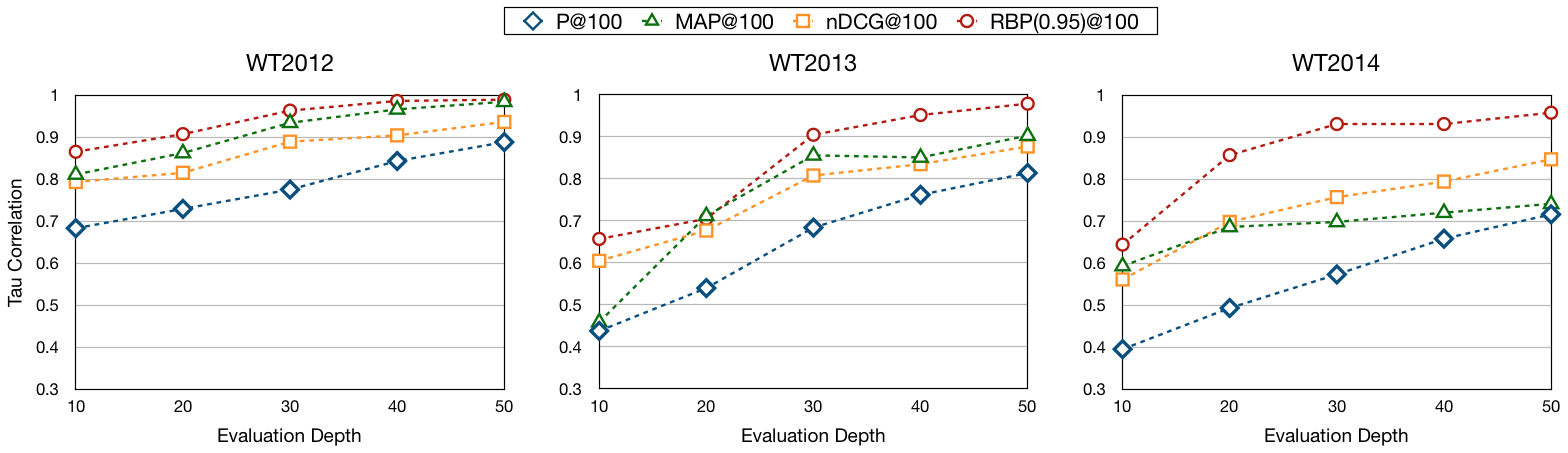}
        \caption{Judgment Depth of High-Cost Measures is 100}
         \label{figure_low_to_high_cost_100}
     \end{subfigure}
     \caption{Prediction of High-Cost Measures Using Low-Cost Measures}
     \label{figure_low_to_high_cost}
\end{figure}

Our prediction results  encouraged us to investigate  whether we could also predict high-cost measures using low-cost measures. We focus on P@1000, P@100, MAP@1000, MAP@100, nDCG@1000, nDCG@100,  RBP@1000, and RBP@100 as the high-cost measures.
As the low-cost measures, we calculate precision, bpref, ERR, infAP\cite{yilmaz2006estimating}, MAP, nDCG and RBP scores of systems when \emph{evaluation depth} (D) is varied from 10 to 50. We specifically use bpref and infAP since they are designed for evaluating systems with incomplete  relevance judgments. We set the $p$ parameter of RBP to 0.95.
For a particular evaluation depth, we calculate  the powerset of the 7 measures mentioned above (excluding the empty set). Subsequently, in a similar approach in Section~\ref{section_predict_with_k_measures}, we find which elements of the powerset are the best predictors of the high-cost measures on WT2012. The set of low-cost measures that yields the maximum $\tau$ score for a particular high-cost measure is also used for predicting the respective measure on WT2013 and WT2014. We repeat this process for each evaluation depth value (i.e. 10, 20, ..., 50)  separately in order to see impact of the cost on the prediction. The results are shown in \textbf{Figure~\ref{figure_low_to_high_cost}}.

For depth 1000 (Figure~\ref{figure_low_to_high_cost_1000}), we achieve higher than 0.9 Kendall's $\tau$ correlation  and higher than 0.98 $R^2$
for RBP in all cases when evaluation depth of low-cost measures is 30 or more. 
While we are able to reach 0.9 $\tau$ correlation for MAP on WT2012, prediction of P@1000 and nDCG@1000 measures performs poorly and never reaches a high $\tau$ correlation. 
As expected, the performance of prediction increases when evaluation depth of high-cost measures are decreased to 100 (Figure~\ref{figure_low_to_high_cost_1000} vs. Figure~\ref{figure_low_to_high_cost_100}). 

Overall, RBP seems the most predictable measure using the low-cost measures while precision is the least predictable one. This is because MAP, nDCG and RBP give more weight to documents at higher ranks, which are also evaluated by the low-cost measures. On the other hand, in calculation of precision, we consider only the number of relevant documents and ignore the ranks.

\section{Conclusion}
\label{section:conclusion}
In this work, we investigated correlation and prediction of evaluation measures using data from 8 TREC test collections covering ad hoc search task for web documents and news articles. 

We first calculated the correlation between 23 evaluation measures. We found that the following measure groups are strongly correlated each other: 1) MAP \& R-Prec \& nDCG, 2) RR \& RBP(0.5), 3) nDCG@20 \& RBP(0.8), 4) P@10 \&  P@20 \& RBP(0.8) \& RBP(0.95). Subsequently, we built a linear regression model to predict a system's evaluation measure using its other  measures and investigated prediction of 12 measures. We found out that we can  predict  MAP, P@10, RBP(0.5) and RBP(0.8) accurately.
Finally, we investigated prediction of high-cost measures using low-cost measures and  showed that we can   predict RBP(0.95) with high accuracy using measures with evaluation depth of 30.





In the future, we plan to deepen our investigation using more data from different tasks and exploring other evaluation metrics and prediction models. 

%

\bibliographystyle{splncs}
{\small
\bibliography{References}{}
}

\end{document}